\documentclass[twocolumn,preprintnumbers,amsmath,amssymb]{revtex4}

\usepackage{graphicx}
\usepackage{dcolumn}
\usepackage{bm}

\begin{document}

\title{Modeling molecular crystals formed by spin-active metal complexes \\ by atom-atom potentials}

\author{Anton V. Sinitskiy}
\email{sinitsk@mail.ru}
\affiliation{Poncelet Laboratory, Independent University of Moscow, Bolshoy Vlasyevskiy Pereulok 11, 119002, Moscow, Russia}

\author{Andrei L. Tchougr\'eeff}
\email{andrei.tchougreeff@ac.rwth-aachen.de}
\affiliation{Poncelet Laboratory, Independent University of Moscow, Bolshoy Vlasyevskiy Pereulok 11, 119002, Moscow, Russia and \\
JARA, Institut f\"ur Anorganische Chemie, RWTH Aachen, Landoltweg 1, 52056 Aachen, Germany}

\author{Andrei M. Tokmachev}
\author{Richard Dronskowski}
\affiliation{JARA, Institut f\"ur Anorganische Chemie, RWTH Aachen, Landoltweg 1, 52056 Aachen, Germany}

\date{\today}

\begin{abstract}

We apply the atom-atom potentials to molecular crystals of iron (II) complexes with bulky organic ligands. The crystals under study are formed by low-spin or high-spin molecules of Fe(phen)$_{2}$(NCS)$_{2}$ (phen = 1,10-phenanthroline), Fe(btz)$_{2}$(NCS)$_{2}$ (btz = 5,5$^{\prime }$,6,6$^{\prime }$-tetrahydro-4\textit{H},4$^{\prime}$\textit{H}-2,2$^{\prime }$-bi-1,3-thiazine), and Fe(bpz)$_{2}$(bipy) (bpz = dihydrobis(1-pyrazolil)borate, and bipy = 2,2$^{\prime }$-bipyridine). All molecular geometries are taken from the X-ray experimental data and assumed to be frozen. The unit cell dimensions and angles, positions of the centers of masses of molecules, and the orientations of molecules corresponding to the minimum energy at 1 atm and 1 GPa are calculated. The optimized crystal structures are in a good agreement with the experimental data. Sources of the residual discrepancies between the calculated and experimental structures are discussed. The intermolecular contributions to the enthalpy of the spin transitions are found to be comparable with its total experimental values. It demonstrates that the method of atom-atom potentials is very useful for modeling organometalic crystals undergoing the spin transitions.

\end{abstract}

\maketitle

\section{\label{intro}Introduction}

The Crystal Field Theory (CFT), proposed in \cite{CFT1} and known to majority of chemists through \cite{CFT2}, suggests that coordination compounds of $d$-elements with electronic configurations $d^{4}$, $d^{5}$, $d^{6}$ or $d^{7}$ can exist either in high-spin (HS) or low spin (LS) forms (sometimes intermediate values of the total spin are also possible). In the case of strong-field ligands the $d$-level splitting measured by the average crystal field parameter $10Dq$ exceeds the average Coulomb interaction energy of $d$-electrons $P$ and the ground state is LS. In the case of weak-field ligands with $10Dq\ll P$, the ground state is bound to be HS. If, however, $10Dq\cong P$, the LS and HS forms of the complex may coexist in equilibrium, and the fraction of either spin form depends on temperature, pressure, and/or other macroscopic thermodynamic parameters. The process when the fraction of molecules of different total spin changes due to external conditions is called a \textit{spin crossover} (SC) transition. For the first time this phenomenon was reported in 1931 \cite{cambi}. Nevertheless, extensive studies of SC started only in 1960s-70s. Nowadays, dozens of complexes capable to undergo spin transitions (spin-active complexes) are known, and most of them are those of Fe(II). A general review of the field can be found in \cite{overview2004}.

A wealth of potential practical applications like displays and data storage devices (see a detailed review in \cite{applic2004}) is one of the reasons for research activity in this area. Industrial applications pose strict demands on the characteristics of the materials to be used. As a consequence, the problem of predicting SC transition characteristics (whether it is smooth or abrupt, the transition temperature, the width of the hysteresis loop, the influence of additives \cite{i2}) is of paramount importance. Theoretical description of spin transitions is a great challenge by itself, and until now a coherent theory allowing to relate the composition of the materials with the characteristics of the transition has not been developed. Discussion of these issues and an overview of the existing theories are given in \cite{NeedForNewModels}.

In general, the SC modeling includes two aspects: (i) that of the interactions within one molecule of a spin-active complex, and (ii) that of the interactions between these molecules. The latter is crucially important for understanding of specific features of the SC transitions in solids because the SC manifesting itself as a first-order phase transition is controlled by intermolecular interactions. These ideas are built in the simplest model capable of describing spin transitions in solids proposed by Slichter and Drickamer \cite{SlDr}. This model considers the solid as a regular solution of molecules in the LS and HS states. The model predicts, in agreement with the experiments, that the spin transition may be either smooth or abrupt or may exhibit hysteresis, and its character is determined by a phenomenological intermolecular parameter $\Gamma $, specific for each material. However, the experimental data on the heat capacity and the X-ray diffraction contradict to this model.

The thermal dependence of the heat capacity of the Fe(phen)$_{2}$(NCS)$_{2}$ crystal is better explained by an alternative domain model \cite{SoraiSeki}. Diffraction patterns of spin transition crystals, measured at intermediate temperatures, simultaneously contain the Bragg peaks corresponding to the pure LS and HS phases, while no peaks for intermediate lattice of a solution were observed \cite{domainsExp}. Another problem is that the parameter $\Gamma $ is phenomenological one, and it cannot be sequentially derived in terms of microscopic characteristics of the constituent molecules or their interactions. At the same time, within the Slichter-Drickamer model, the type of behavior is tightly related to the sign and magnitude of $\Gamma $, so that a smooth transition requires $\Gamma >0$, an abrupt transition occurs at $\Gamma <0$ and hysteresis is possible only if $\Gamma <0$ is less than some critical threshold, which in its turn depends on the transition temperature \cite{SlDr}. It has been shown that if the relaxation of the lattice is not allowed, then under very natural assumptions $\Gamma $ is positive \cite{i11}, but the lattice relaxation can lead to $\Gamma $ of either sign \cite{TDIJQC}.

Significant progress in the understanding of the spin transitions in crystals is attributed to the Ising-like models of intermolecular interactions in spin-active materials \cite{BolvinKahn1995}. Adaptations of the initial Ising model to the spin transitions include corrections for intramolecular vibrations, domain formation, parameters distribution, elastic distortions, presence of two metal atoms in a spin-active molecule, \textit{etc.} \cite{NeedForNewModels,Ising}. These models do not have analytical solutions and they are solved either in a mean field approximation which leads to results analogous to (or even coinciding with) the Slichter-Drickamer model \cite{NeedForNewModels} or numerically.
%and references therein; also recent works \cite{Jap2008,
%Jap2007, 2step1997} and argumentation for including magnetic exchange
%between the HS Fe(II) centers into the Hamiltonian together with the elastic
%interactions in \cite{Jeschke2007}). (Lattice distortions stemming from
%the molecular interactions at a microscopic level may also be discussed
%without any reference to the Ising model, like in \cite{TDIJQC, Spiering2004}
%.)

In spite of the diversity of the models used in the literature, the theoretical description of the spin transitions is not yet satisfactory. First, the existing theories are not capable to reproduce the whole set of the experimental data (\textit{e.g.} asymmetry of the hysteresis loop \cite{NeedForNewModels}). Second, all of them contain phenomenological parameters, like $\Gamma $ in the Slichter-Drickamer model, or the energy gap $\Delta _{i}$ or the interspin interaction constants $J_{ij}$ in the Ising-like models, or the bulk modulus $K$ and the Poisson ratio $\sigma $ in \cite{Spiering2004} ($K$ and $\sigma $ can be measured, but for the purpose of the theory they must be independently predicted), \textit{etc.} Third, even if the models include microscopic level consideration, they use oversimplified description of the molecules (as spheres, ellipsoids), which is not sufficient for constructing a complete theory, especially due to importance of the short intermolecular contacts tentatively responsible for the cooperativity effects ($\pi $-$\pi $ interactions, S$\cdots $H$-$C interactions, hydrogen bonds, \textit{etc.} \cite{FeNCS2004}).

These shortcomings can be overcome by using explicit potentials for intra- and intermolecular interactions. In this case one may expect to obtain independent estimates of the numerous parameters required by the phenomenological theories. These potentials should also be a helpful tool for checking the validity of the initial postulates, such as the formation of a regular solution or the domain structure, thus clarifying some obscure points in the theory itself.

An adequate \textit{ab initio} calculation of the energies of isolated transition metal complexes, and moreover those of the crystals formed by these complexes, is a very complicated problem. Significant electron correlation within the $d$-shells breaks the self-consistent field approximation, so that explicit account of nontrivial (static) electron correlation is unavoidable. The existing implementations of \textit{ab initio} approaches for solids fail to provide the necessary quality of the results.

There is a number of attempts to use the DFT-based methods to take into account the electron correlation in the SC complexes \cite{DFToverview2004}. These methods yield good results for many characteristics of isolated spin-active molecules (optimal molecular geometry, M\"ossbauer parameters, vibrational frequencies, nuclear inelastic scattering spectra) \cite{dftmol,borshch2005,lemercier2006,letard2008}. However, the DFT in its traditional form, as it is demonstrated in \cite{Tch075}, is not capable to reproduce coherently the static correlations, which are extremely important for the correct description of the spin transitions even in an isolated molecule. For that reason the results for the energy gap between the LS and HS states, and hence for the transition temperature, obtained by the DFT techniques are absolutely disastrous. The common versions of DFT, such as B3LYP, often predict a wrong ground state multiplicity, let alone the value of the energy difference \cite{DFToverview2004}. For example, the temperature of the spin transition in Fe(phen)$_{2}$(NCS)$_{2}$ was found to be an order of magnitude too large (1530 K instead of 176 K) \cite{reicher2002}. In addition, most DFT studies are limited to isolated molecules \textit{in vacuo}, and the heat of the spin transitions in a crystal is identified with the energy difference of isolated molecules. The influence of the intermolecular interactions is thus neglected.

Only a few isolated attempts to explicitly model a spin-active crystal by the DFT method have been reported \cite{lemercier2006,letard2008,Jeschke2007,Angyan2008}. The application of the LDA approximation with the periodic boundary conditions to the crystals of Fe(trim)$_{2}$X$_{2}$ (X = F, Cl, Br, or I, and trim = 4-(4-Imidazolylmethyl)-2-(2-imidazolylmethyl)imidazole) formed by either LS or HS molecules \cite{lemercier2006} demonstrated that the intermolecular interactions strongly affect the energy splitting between the LS and HS isomers, thus necessitating their adequate treatment within coherent SC models. The experimental X-ray structures for some complexes are available, so that the optimal geometry of the crystals found by LDA can be verified. This comparison showed that the unit cell volumes were overestimated by 20-24\%. At the same time, the calculated N...X distances were 0.1-0.3 \AA \ lower than the experimental ones, and the $\pi $-stacking distances were underestimated by 0.3-0.7 \AA . Although in general it is difficult to separate the errors from intra- and intermolecular interactions, the geometry of individual spin isomers is usually described much better than the relative position of the molecules in the crystal. The GGA approximation has been used to optimize molecular geometry and lattice parameters of the LS and HS crystals of [Fe(pyim)$_{2}$(bipy)](ClO$_{4}$)$_{2}\cdot $2C$_{2}$H$_{5}$OH (pyim = 2-(2-pyridyl)imidazole) \cite{Postnikov2008}. The bond lengths were found to be quite reasonable. However, the lattice parameters were poorly reproduced, so that even the wrong sign of the unit cell volume change for the LS to HS transition was obtained: ($-$6.82 \AA $^{3}$ instead of experimental value of +228.02 \AA $^{3}$).
The authors explain it by \textquotedblleft the well-known shortcomings of DFT methods in application
to weak intermolecular interactions\textquotedblright \ \cite{Postnikov2008}. The DFT+$U$ approach
with the GGA approximation has been applied to model spin-active crystals of Fe(phen)$_{2}$(NCS)$_{2}$
and Fe(btr)$_{2}$(NCS)$_{2}$(H$_{2}$O) (btr = 4,4$^{\prime }$-bis-1,2,4-triazole) \cite{Angyan2008}.
The studies of the Fe(phen)$_{2}$(NCS)$_{2}$ crystal
have demonstrated that DFT is capable of reproducing the lattice parameters with the precision of 1-5\%
 and the unit cell volume with the precision of $2-7$\% \cite{AngyanPrivComm}.
Unfortunately, these works do not employ a much better substantiated
approach to the DFT-based treatment of van der Waals interactions previously proposed
by the same authors, based on
explicit treatment of correlations coming from the long range part of the
electron-electron interactions \cite{GerberAngyan}.
% However,
%neither of these methods have been applied to describing the spin-active
%materials.

Summarizing, DFT models either produce poor results for spin-active complexes or require parameters (like DFT+$U$) adjusted to reproduce the experimental data. At the same time, the very idea of modeling such complex system as a crystal formed by spin-active transition metal complexes at a uniform level of theory seems to be incorrect. The systems under consideration consist of numerous components, and it is much more natural to treat these components separately -- each at the adequate level of the theory. The most important separation is that on intra- and intermolecular interactions. On the level of molecules one can further separate a highly correlated $d$-shell from the rest of the molecule. This idea has been implemented as a specialized quantum chemical method -- Effective Hamiltonian of Crystal Field (EHCF) \cite{i22} which has been successfully applied to describe the spin isomers of Fe(phen)$_{2}$(NCS)$_{2}$ \cite{i36}. Furthermore, it has been demonstrated that the geometry of spin-active complexes can be adequately described by the EHCF technique with ligands treated by molecular mechanics force fields \cite{Tch065}.
%The nature of the object of our study --
%molecular crystals formed by spin-active complexes -- allows us to simplify
%the solution yet further.

On the level of interactions between molecules the paramount fact is that the molecular crystals formed by spin-active molecules consist of complexes with bulky organic ligands. Intermolecular contacts in such crystals are those between the organogenic atoms like C, H, N, S, \textit{etc.} The $d$-shells of the central ions are effectively shielded by the ligands. Thus, it is reasonable to assume \cite{i11} that the $d$-shells do not directly affect the interactions between the molecules of the different total spin in the crystal, but influence it indirectly: through the variation of the equilibrium interatomic distances Fe$-$N in these complexes, which is further translated into different "sizes" of the LS and HS isomers. In this context, the standard methods developed for organic molecular crystals can be successfully applied in this case as well. The main purpose of the present work is to identify an adequate way to model intermolecular interactions for crystals formed by spin-active molecules.

\section{\label{poten}Atom-atom potentials method}

In order to avoid unnecessary complications, we limit our task in the present paper to checking the possibility of applying the simplest method of modeling intermolecular interactions -- atom-atom potentials \cite{i24} -- to crystals formed by spin-active complexes. The method
%is based on the following assumptions:
%
%\begin{enumerate}
%\item The energy of the intermolecular interaction equals to the sum of the energies of interactions between the single atoms forming the molecules (additivity).
%
%\item The forces between the atoms are central, \textit{i.e.} depend only on the distances between them (spherical symmetry).
%
%\item The energy of the interaction between two atoms is a sum of dispersion, repulsion, and Coulomb components. (The Coulomb interactions are omitted from the current treatment.)
%
%\item The energy of the interatomic interaction depends on the type of atoms only, not on their valence state or molecular environment (transferability or universality of the parameters).
%\end{enumerate}
%
%As a result, 
assumes that the energy of the molecular crystal (calculated relative to the system of isolated molecules) can be represented as:
\begin{equation}
U=\frac{1}{2}\sum _{\alpha \alpha ^{\prime }mm^{\prime }\overline{r}\overline{r}^{\prime }}E_{\alpha \alpha ^{\prime }}\left( R\left( \alpha \alpha ^{\prime }mm^{\prime }\overline{r}\overline{r}^{\prime }\right) \right),
\label{uuu}
\end{equation}
where each term is the energy of the interaction between the $\alpha $-th atom of the $m$-th molecule in the unit cell number $\overline{r}=(r_{a},r_{b},r_{c})$ and the $\alpha ^{\prime }$-th atom of the $m^{\prime }$-th molecule in the unit cell $\overline{r}^{\prime }$ depending on the distance $R$. Due to the equivalence of all unit cells, we can get rid of summation over $\overline{r}^{\prime }$, and the energy per molecule $u$ can be written as:
\begin{equation}
u=\frac{1}{2M}\sum _{\alpha \alpha ^{\prime }mm^{\prime }\overline{r}}E_{\alpha \alpha ^{\prime }}\left( R\left( \alpha \alpha ^{\prime }mm^{\prime }\overline{r}0\right) \right),
\label{uuu1}
\end{equation}
where $M$ is the number of molecules per unit cell.

A number of approximations have been suggested for the atom-atom interaction. The most widespread ones are the Buckingham potential (6-$exp$):
\begin{equation}
E_{\alpha \alpha ^{\prime }}(R)=-\frac{A_{\alpha \alpha ^{\prime }}}{R^{6}}+B_{\alpha \alpha ^{\prime }}e^{-C_{\alpha \alpha ^{\prime }}R},
\label{Buck}
\end{equation}
and the Lennard-Jones potential (6-$n$):
\begin{equation}
E_{\alpha \alpha ^{\prime }}(R)=-\frac{A_{\alpha \alpha ^{\prime }}}{R^{6}}+\frac{B_{\alpha \alpha ^{\prime }}}{R^{n}}.
\label{LJ}
\end{equation}
In the above formulae the $A_{\alpha \alpha ^{\prime }}$ and $B_{\alpha \alpha ^{\prime }}$ parameters for the interaction between atoms of different types are often calculated as the geometric mean values of the corresponding homogeneous interaction parameters:
\begin{equation}
A_{\alpha \alpha ^{\prime }}=\sqrt{A_{\alpha \alpha }A_{\alpha ^{\prime }\alpha ^{\prime }}},\quad B_{\alpha \alpha ^{\prime }}=\sqrt{B_{\alpha \alpha }B_{\alpha ^{\prime }\alpha ^{\prime }}}, \label{superposAB}
\end{equation}
while the $C_{\alpha \alpha ^{\prime }}$ parameter is approximated in a similar way as an arithmetic mean value:
\begin{equation}
C_{\alpha \alpha ^{\prime }}=\frac{1}{2}(C_{\alpha \alpha }+C_{\alpha ^{\prime }\alpha ^{\prime }}).
\label{superposC}
\end{equation}

Due to these approximations, the energy can be represented as a fast computable function depending on the lattice parameters and relative positions and orientations of the molecules in the unit cell provided that molecular geometry of the complex is fixed. Having found the minimum of this function, one gets estimates of the intermolecular interaction energy (sublimation energy), the equilibrium unit cell parameters, and the positions and orientation of the molecules in the unit cell at the absolute zero temperature and absence of external pressure.

One can easily extend the method to account for the external pressure. For this purpose one should optimize the enthalpy $H$ instead of the potential energy $U$. The enthalpy is defined as
\begin{equation}
H=U+PV,
\label{enthalpy}
\end{equation}
where the volume $V$ is determined by the lattice parameters. As for the thermal dependence of the lattice parameters, the matter is not so simple. One should basically minimize \textit{the Gibbs energy} $G$ to estimate the equilibrium values of the lattice parameters at a non-zero temperature (and pressure). This procedure includes calculation of the entropy of the crystal undergoing the spin transition, which is a separate non-trivial challenge, as shown in \cite{entropy}. To avoid this, one may confine to minimization of the internal energy $U$ or the enthalpy $H$, but the resulting lattice parameters will be relevant only for the absolute zero of temperature. On the other hand, in practice the parameters of atom-atom interaction are fitted in such a way that the lattice parameters corresponding to the minimum of the model \textit{internal energy} $U$ best reproduce the experimental lattice structures measured at \textit{the room temperature} (see \textit{e.g.} \cite{Buckparam}). In this case the model includes the entropy factor implicitly, and the lattice parameters found by direct minimization of $U$ should actually refer to the room temperature.

The accuracy of the atom-atom approach is corroborated by extensive statistics obtained for organic molecular crystals \cite{i24,gavezzotti1994,price2008}. Typically it provides the accuracy level of \textit{ca.} 0.1$\div $5 kcal/mol in energy terms for a wide range of organic crystals. However, in theory we can expect much better precision for the \textit{relative} energies of the crystals undergoing the spin transition, since the LS and HS crystals are very similar to each other (as is shown below, the shortest contacts are the same).

\section{\label{method}Modeling method}

\begin{figure}
\includegraphics[scale=0.4]{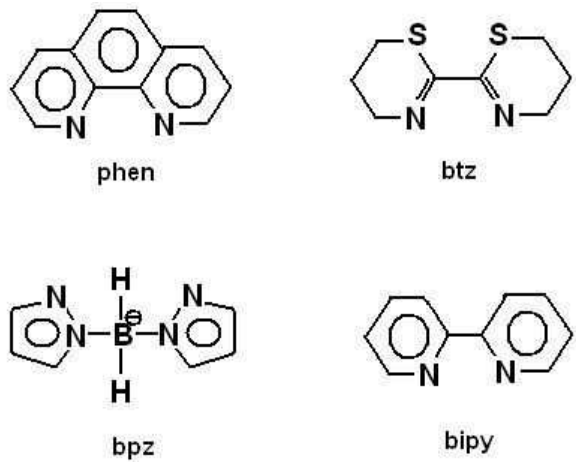}
\caption{\label{ligands}Structure formulae for the ligands of the spin-active complexes studied.}
\end{figure}

\begin{figure}
\includegraphics[scale=0.4]{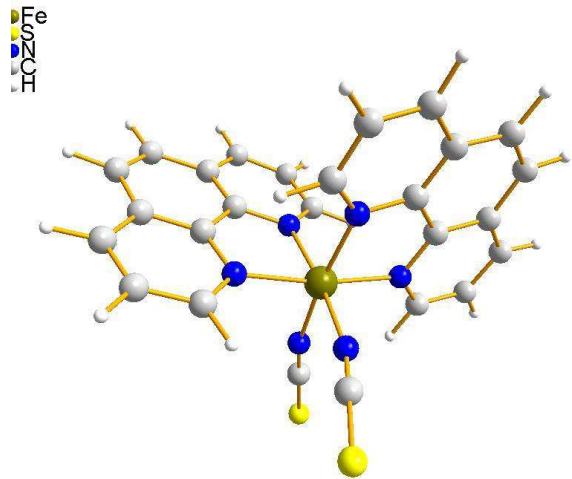}
\caption{\label{phen_mol}Molecular structure of Fe(phen)$_{2}$(NCS)$_{2}$.}
\end{figure}

\begin{figure}
\includegraphics[scale=0.4]{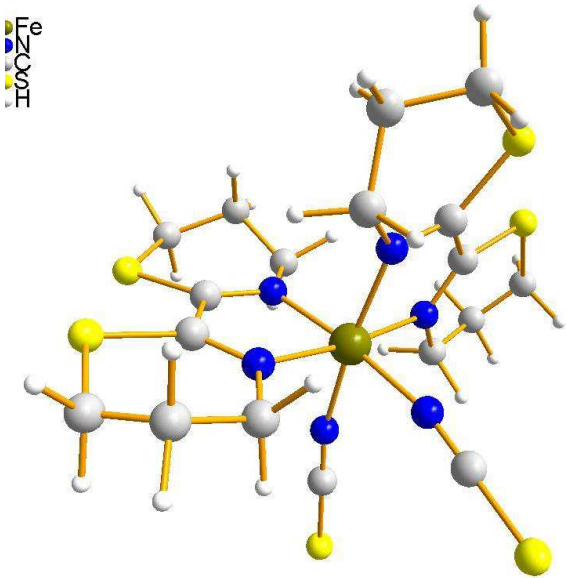}
\caption{\label{btz_mol}Molecular structure of Fe(btz)$_{2}$(NCS)$_{2}$.}
\end{figure}

\begin{figure}
\includegraphics[scale=0.4]{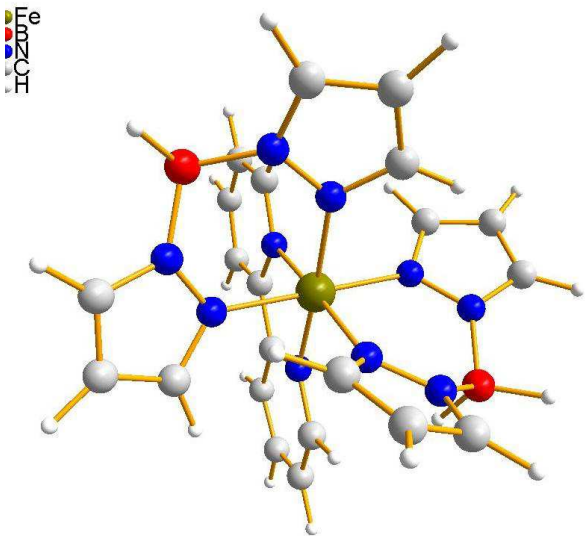}
\caption{\label{bpz_mol}Molecular structure of Fe(bpz)$_{2}$(bipy).}
\end{figure}

We performed calculations for the molecular crystals formed by each of the spin isomers of Fe(phen)$_{2}$(NCS)$_{2}$, Fe(btz)$_{2}$(NCS)$_{2}$, and Fe(bpz)$_{2}$(bipy). The ligands are depicted in Fig. \ref{ligands} and the molecules themselves are shown in Figs. \ref{phen_mol}-\ref{bpz_mol}. The objects were chosen based on the following considerations. First, all these crystals consist of neutral molecules only, without ions or solvents. As a result, the molecules are held together in the crystal by the van der Waals forces (no strong Coulomb forces or obvious hydrogen bonds are involved), which dramatically simplifies modeling of the energy. Second, these three substances represent all main types of spin transitions: abrupt one in the Fe(phen)$_{2}$(NCS)$_{2}$ crystal, smooth one in the Fe(btz)$_{2}$(NCS)$_{2}$ crystal, and the transition with hysteresis in the Fe(bpz)$_{2}$(bipy) crystal. Finally, the crystallographic data (including the molecular geometries) for both HS and LS forms of these three substances are available in the literature.
%where phen stands for
%1,10-phenanthroline; Fe(btz)$_{2}$(NCS)$_{2}$, where btz =
%5,5$^{\prime}$,6,6$^{\prime}$-tetrahydro-4{\it H},4$^{\prime}${\it H}-2,2$^{\prime}$-
%bi-1,3-thiazine;
%Fe(bpz)$_{2}$(bipy), where bpz =
%dihydrobis(1-pyrazolil)borate, bipy = 2,2$^{\prime}$-bipyridine, 
%with the ligands depicted in Fig. 1%\ref{ligands}
%\footnote{Note: the original 
%paper \cite{BtzPhenExp} includes an incorrect name for btz --
%2,2$^{\prime}$-bis-4,5-dihydrothiazine -- and does not provide its structural 
%formula.
%Authors of the review \cite{FeNCS2004} attempted to correct the name and 
%provide the structural
%formula, but their version is incorrect as well, as the NMR spectra and
%the synthetic procedure description \cite{BtzPhenExp} consideration shows.
%}

The energy of van der Waals interactions was described by the Lennard-Jones (6-12) potential with the parameters of the "Universal Force Field" (UFF) parameterization \cite{UFFparam} and by the Buckingham (6-$exp$) potential with the parameters provided in \cite{Buckparam} (see Tables \ref{LJparams}, \ref{Buckparams}). In the latter case the parameters for the C$\cdots $H, N$\cdots $H, S$\cdots $H, C$\cdots $N, and S$\cdots $C interactions are given in \cite{Buckparam} explicitly and there is no need to use eqs. (\ref{superposAB}) and (\ref{superposC}). Unfortunately, the system of parameters \cite{Buckparam} for the (6-$exp$) potential has not been extended to boron. Hence we took the minimum depth and the interatomic separation at the minimum for the B$\cdots $B pair from \cite{Pletnev}, estimated the corresponding $A$, $B$ and $C$ parameters and found the parameters for the B$\cdots $H, B$\cdots $C, B$\cdots $N, and B$\cdots $S pairs following eqs. (\ref{superposAB}) and (\ref{superposC}). The parameters of the (6-$exp$) potential for pairs involving Fe atom(s) are not determined, but they are immaterial in the present context, and we set them to be equal to zero.

\begin{table}
\caption{\label{LJparams}Parameters of the Lennard-Jones (6-12) potential \cite{UFFparam} used in the calculations.}
\begin{ruledtabular}
\begin{tabular}{|l|c|c|c|c|c|c|}
\hline
& H & B & C & N & S & Fe \\ \hline
A, kcal$\cdot $\AA$^{6}$/mol & 50.9 & 1668 & 685 & 332 & 2365 & 15.9 \\ \hline
B, 10$^{7}\cdot $kcal$\cdot $\AA$^{12}$/mol & 0.147 & 38.6 & 11.2 & 3.99 & 51 & 0.048 \\ \hline
\end{tabular}
\end{ruledtabular}
\end{table}

\begin{table*}
\caption{\label{Buckparams}Parameters of the Buckingham (6-$exp$) potential \cite{Buckparam} used in the calculations.}
\begin{ruledtabular}
\begin{tabular}{|l|c|c|c|c|c|c|c|c|c|c|}
\hline
& H$\cdots $H & C$\cdots $H & N$\cdots $H & S$\cdots $H & C$\cdots $C & C$\cdots $N & S$\cdots $C & N$\cdots $N & S$\cdots $S & B$\cdots $B \cite{Pletnev} \\ \hline
A, kcal$\cdot $\AA$^{6}$/mol & 26.1 & 113 & 120 & 279 & 578 & 667 & 1504 & 691 & 2571 & 3.688 \\ \hline
B, 10$^{3}\cdot $kcal/mol & 5.774 & 28.87 & 54.56 & 64.19 & 54.05 & 117.47 & 126.46 & 87.3 & 259.96 & 19.84 \\ \hline
C, \AA$^{-1}$ & 4.01 & 4.10 & 4.52 & 4.03 & 3.47 & 3.86 & 3.41 & 3.65 & 3.52 & 6.82 \\ \hline
\end{tabular}
\end{ruledtabular}
\footnotetext{Note: parameters for the S$\cdots $N interaction and interactions of B with other atoms were calculated according to the superposition approximation (\ref{superposAB}) and (\ref{superposC}).}
\end{table*}

The MOLCRYST program suite \cite{i25} capable of calculation and minimization of molecular crystals energy and enthalpy with use of the Lennard-Jones and Buckingham atom-atom potentials was employed. This program has been thoroughly tested on the examples of molecular crystals of aromatic hydrocarbons. The geometries of HS and LS forms of the complexes were taken from experiments \cite{PhenExp,BtzPhenExp,BpzExp} and were assumed to be fixed (frozen) throughout the modeling. The validity of the rigid-body approximation can be tested \cite{trueblood} and the analysis of the difference vibrational parameters for an SC crystal demonstrated \cite{buergi} that the non-rigidity is relatively small for both HS and LS forms.

When calculating the energy according to eq. (\ref{uuu1}), we restricted ourselves to summation over three layers of unit cells around the central "0-th" unit cell. In other words, only those $\overline{r}=(r_{a},r_{b},r_{c})$ were included into the sum, for which $\left\vert r_{a}\right\vert \leq 3$, $\left\vert r_{b}\right\vert \leq 3$, and $\left\vert r_{c}\right\vert \leq 3$. It was found that extending this limit to 4 or more layers does not affect the final result for the energy or enthalpy (the differences are less than $0.01$ kcal/mol). As for the equilibrium values of the lattice parameters, their values are stable (within $0.1$\%) already with one layer of the surrounding unit cells (those adjacent to the "0-th" cell) included into the summation.

\begin{figure}
\includegraphics[scale=0.44]{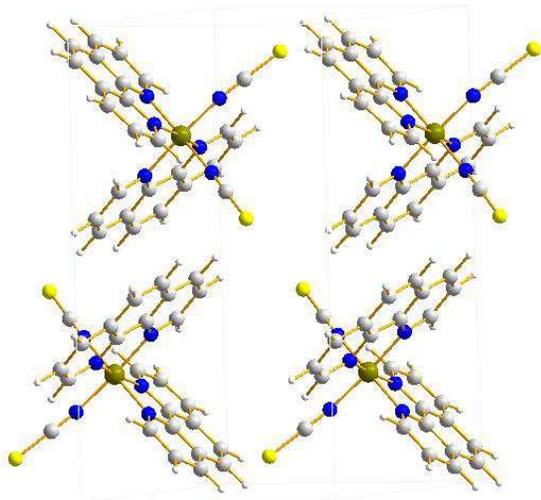}
\caption{\label{phen_str}Crystal structure of Fe(phen)$_{2}$(NCS)$_{2}$.}
\end{figure}

\begin{figure}
\includegraphics[scale=0.44]{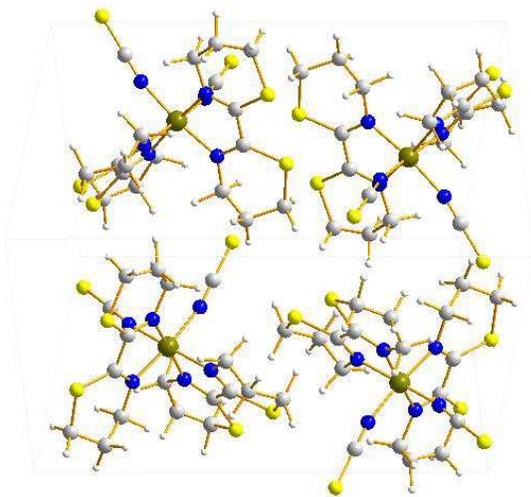}
\caption{\label{btz_str}Crystal structure of Fe(btz)$_{2}$(NCS)$_{2}$.}
\end{figure}

\begin{figure}
\includegraphics[scale=0.44]{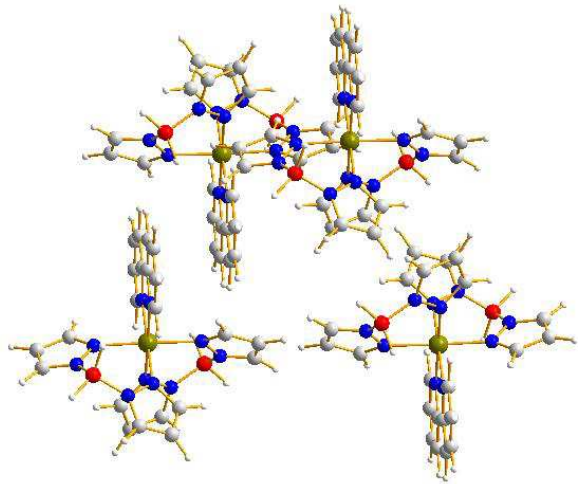}
\caption{\label{bpz_str}Crystal structure of Fe(bpz)$_{2}$(bipy).}
\end{figure}

To find the equilibrium values of the lattice parameters, positions of the centers of masses (CM), and the rotation angles of molecules in the unit cell, minimization of the \textit{enthalpies} of six pure crystals (three HS and three LS) was performed. Pressure was set to be 1 atm. In all the cases the experimental crystallographic data were taken as initial approximations. At the first stage we minimized the enthalpy as a function of five or six parameters ($a$, $b$, $c$, one non-trivial rotation angle, and one non-trivial CM coordinate in the cases of the Fe(phen)$_{2}$(NCS)$_{2}$ and Fe(btz)$_{2}$(NCS)$_{2}$ crystals; the same plus the unit cell angle $\beta$ in the case of the Fe(bpz)$_{2}$(bipy) crystals), preserving the symmetry of the crystal (Pbcn, Pbcn and C2/c correspondingly); after that we checked that the final point of the previous step is the global minimum, allowing for variation of all 27 parameters ($a$, $b$, $c$, three unit cell angles, three rotation angles for each of four molecules in the unit cell, three CM position coordinates for three out of four molecules in the unit cell; the fourth molecule position is not independent due to the crystal translational symmetry). The optimized structures are shown on Figs. \ref{phen_str}-\ref{bpz_str}.

The enthalpy was calculated as the sum of the internal energy and the product of the pressure and the volume of the crystal with 1 mole of molecules. In all the cases the internal energies were found to be about $-50$ kcal/mol relative to the isolated molecules. The experimental data to verify this result are not available. However, the energy magnitude is quite reasonable in comparison with the available data on organic molecular crystals \cite{gavezzotti1994,gavezzotti1999}, taking into consideration that the numbers of interatomic contacts per molecule in the crystals under study are a few times higher than those in ordinary organic crystals. The differences between the internal energies and the enthalpies in all the cases at 1 atm are rather small, less than $0.01$ kcal/mol, which is not surprising, since we deal with solid substances.
%Despite that extensive set of optimized
%variables, the whole calculation (energy minimization) for one substance at
%a specific pressure takes less than one minute on a PC, which is absolutely
%unattainable with use of the DFT-based methods.

As mentioned above, the (6-$exp$) potential parameterization from \cite{Buckparam} implicitly includes the entropy contribution since it was fitted to reproduce the room temperature geometries of crystals by minimization of the internal energy rather than the Gibbs energy. So the results of our calculation with the Buckingham potential should be compared with the room temperature experimental data. The matter is not so clear in the case of the UFF parameter system \cite{UFFparam}. The authors introduce their parameters of the van der Waals interaction explicitly referring to ionization potentials, polarizabilities, and Hartree-Fock calculations, so these parameters seem to be providing unadjusted estimates of the internal energy. Nevertheless, direct comparisons of the numbers produced with their empirical parameters and experimental geometries are widely used. Strictly speaking, we do not have sufficient information to judge whether our results obtained with this (6-12) potential describe physical properties for the absolute zero temperature or for the room temperature. However, comparing the experimental data on the lattice parameters of Fe(phen)$_{2}$(NCS)$_{2}$ at 15 K, 32 K, 130 K and 298 K \cite{phenLowT} with the results of our calculations, we can see that the latter are somewhat closer to the high-temperature values of the lattice parameters, rather than to the low-temperature ones.

The room temperature crystallographic data are available only for the HS crystals. As for the LS crystals, we need to extrapolate their experimental lattice parameters to the room temperatures to make the comparison with the results of our calculations possible. This is especially important for the analysis of the changes of the lattice parameters $\Delta V$, $\Delta a$, $\Delta b$, $\Delta c$, \textit{etc.} in the course of the spin transition, otherwise the calculated experimental values would include not only the contribution of the spin transition itself, but also of thermal expansion of the crystal. In the cases of the Fe(phen)$_{2}$(NCS)$_{2}$ and Fe(btz)$_{2}$(NCS)$_{2}$ compounds, the dependences of the $V$, $a$, $b$, $c$ parameters and the HS molecules fraction $x$ (from the magnetic susceptibility data) on temperature are known in the range from \textit{ca.} 130 K to 293 K \cite{BtzPhenExp} (each series consists of 22-25 observations). In a linear approximation,
\begin{equation}
\begin{array}{c}
V\left( T,x(T)\right) =(1-x)\left( V_{o,LS}+\kappa _{V,LS}(T-T_{o})\right) \\
+x\left( V_{o,HS}+\kappa _{V,HS}(T-T_{o})\right) ,
\end{array}
\label{linV}
\end{equation}
and similarly for the $a$, $b$ and $c$ parameters. We determined the coefficients $V_{o,LS}$, $\kappa _{V,LS}$, $V_{o,HS}$, $\kappa _{V,HS}$, \textit{etc.} by the method of least squares ($R^2$ of such models are typically 0.995$\div $0.999), and made extrapolation of the lattice parameters to the room temperature and unchanged fraction of the HS molecules. These extrapolated values were used for comparison with the results of the method of atom-atom potentials. As for the Fe(bpz)$_{2}$(bipy) crystal, the lattice parameters, published in the literature, were measured only at few temperatures \cite{BpzExp, BpzLowT}. Thermal coefficients of expansion, calculated for the LS form on different temperature intervals, differ significantly, which does not allow for a reliable extrapolation of the lattice parameters to the room temperature. On the other hand, high-temperature coefficients of expansion are more stable. Because of these reasons, we extrapolated the HS crystal lattice parameters to 139 K to estimate $\Delta V$, $\Delta a$, $\Delta b$ and $\Delta c$ free of thermal distortions, though only at 139 K. The results of the extrapolations made are used in the next Section for comparison with calculated optimal lattice parameters.

\section{\label{RD}Results and discussion}

\subsection{\label{geom}Crystal geometries}

The most important for thermodynamical description of the spin transitions characteristic of the lattice is \textit{the unit cell volume}. The estimates of this quantity obtained by the atom-atom potentials model are given in Tables \ref{phenExpCalc}-\ref{bpzExpCalc}. The average error in the computed volume is $1.8$\%, ranging from $0.5$\% to $4.0$\%. The Lennard-Jones and the Buckingham potentials provide comparable levels of accuracy. These numbers should be compared with the discrepancy of $20-24$\% in \cite{lemercier2006} and $1-8$\% in \cite{Postnikov2008} (both calculated with the DFT method), the only analogues published so far. At the same time one should remember that these data include relatively small errors in the geometries of separate molecules while our calculations are free of them because we used the experimental structures for the molecules.
%which is equal to molar volume
%divided by the Avogadro number and multiplied by the number of molecules per
%unit cell.

\begin{table*}
\caption{\label{phenExpCalc}Comparison of experimental and calculated unit cell parameters for Fe(phen)$_{2}$(NCS)$_{2}$ (at 1 atm).}
\begin{ruledtabular}
\begin{tabular}{|l|c|c|c|c|c|c|c|c|}
\hline
& $a$, \AA & $b$, \AA & $c$, \AA & $\beta ,^{\circ }$ & $V$, \AA$^{3}$ & angle,$^{\circ }$ & CM $y/b$ & H, kcal/mol \\ \hline
\multicolumn{9}{|l|}{\textbf{The LS isomer}} \\ \hline
calc. (6-12) & 13.185 & 9.922 & 17.347 & 90 & 2269.2 & 142.85 & 0.1112 & -54.46 \\
calc. (6-$exp$) & 12.992 & 9.861 & 17.281 & 90 & 2214.0 & 144.48 & 0.1138 & -54.38 \\
calc. (6-$exp$) modif. & 13.017 & 9.991 & 17.469 & 90 & 2271.7 & 144.59 & 0.1065 & -54.04 \\
exp. 15 K \cite{phenLowT} & 12.762 & 10.024 & 17.090 & 90 & 2186.3 & 143.84 & 0.0943 & - \\
exp. 130 K \cite{PhenExp} & 12.770 & 10.090 & 17.222 & 90 & 2219.1 & 140.51 & 0.0925 & - \\
exp. extrap. to 293 K & 12.77 & 10.18 & 17.40 & 90 & 2259 & - & - & - \\ \hline
\multicolumn{9}{|l|}{\textbf{The HS isomer}} \\ \hline
calc. (6-12) & 13.525 & 9.910 & 17.583 & 90 & 2356.7 & 147.36 & 0.1071 & -52.65 \\
calc. (6-$exp$) & 13.264 & 9.869 & 17.542 & 90 & 2296.2 & 149.35 & 0.1088 & -53.80 \\
calc. (6-$exp$) modif. & 13.227 & 10.017 & 17.815 & 90 & 2360.4 & 149.19 & 0.0993 & -52.50 \\
exp. 15 K \cite{phenLowT} & 13.185 & 9.948 & 17.135 & 90 & 2247.5 & 153.84 & 0.0989 & - \\
exp. 293 K \cite{PhenExp} & 13.161 & 10.163 & 17.481 & 90 & 2338.2 & 147.09 & 0.0938 & - \\ \hline
\multicolumn{9}{|l|}{\textbf{The difference between the HS and LS isomers}} \\ \hline
calc. (6-12) & 0.340 & -0.012 & 0.236 & 0 & 87.5 & 4.51 & -0.0040 & 1.81 \\
calc. (6-$exp$) & 0.272 & 0.008 & 0.260 & 0 & 82.2 & 4.87 & -0.0050 & 0.57 \\
calc. (6-$exp$) modif. & 0.210 & 0.026 & 0.347 & 0 & 88.7 & 4.60 & -0.0072 & 1.54 \\
exp. extrap. to 293 K & 0.39 & -(0.02$\div $0.04)  & 0.05$\div $0.08 & 0 & 70$\div $79 & - & - & - \\
exp. (15 K) \cite{phenLowT} & 0.423 & -0.076 & 0.045 & 0 & 61.2 & 10.00 & 0.0045 & - \\ \hline
\end{tabular}
\end{ruledtabular}
\footnotetext{Note: see detailed explanation of "(6-$exp$) modif." parameterization in Subsection (\ref{cont}).}
\end{table*}

\begin{table*}
\caption{\label{btzExpCalc}Comparison of experimental and calculated unit cell parameters for Fe(btz)$_{2}$(NCS)$_{2}$ (at 1 atm).}
\begin{ruledtabular}
\begin{tabular}{|l|c|c|c|c|c|c|c|c|}
\hline
& $a$, \AA & $b$, \AA & $c$, \AA & $\beta ,^{\circ }$ & $V$, \AA$^{3}$ & angle,$^{\circ }$ & CM $y/b$ & H, kcal/mol \\ \hline
\multicolumn{9}{|l|}{\textbf{The LS isomer}} \\ \hline
calc. (6-12) & 13.266 & 10.518 & 16.889 & 90 & 2356.4 & 125.60 & 0.0385 & -54.15 \\
calc. (6-$exp$) & 13.099 & 10.498 & 16.741 & 90 & 2302.1 & 126.45 & 0.0445 & -57.10 \\
calc. (6-$exp$) modif. & 13.160 & 10.652 & 16.963 & 90 & 2377.9 & 127.77 & 0.0493 & -52.55 \\
exp. (130 K) \cite{BtzPhenExp} & 13.055 & 10.650 & 16.672 & 90 & 2318.1 & 127.48 & 0.0421 & - \\
exp. extrap. to 293 K & 13.17 & 10.80 & 16.88 & 90 & 2397 & - & - & - \\ \hline
\multicolumn{9}{|l|}{\textbf{The HS isomer}} \\ \hline
calc. (6-12) & 13.242 & 10.724 & 16.947 & 90 & 2406.6 & 129.45 & 0.0451 & -54.32 \\
calc. (6-$exp$) & 13.077 & 10.669 & 16.803 & 90 & 2344.3 & 130.14 & 0.0498 & -58.20 \\
calc. (6-$exp$) modif. & 13.190 & 10.786 & 16.973 & 90 & 2414.5 & 130.77 & 0.0527 & -53.61 \\
exp. (293 K) \cite{BtzPhenExp} & 13.288 & 10.861 & 16.920 & 90 & 2441.9& 129.79 & 0.04150 & - \\ \hline
\multicolumn{9}{|l|}{\textbf{The difference between the HS and LS isomers}} \\ \hline
calc. (6-12) & -0.023 & 0.206 & 0.059 & 0 & 50.2 & 3.85 & 0.0066 & -0.17 \\
calc. (6-$exp$) & -0.022 & 0.172 & 0.062 & 0 & 42.2 & 3.68 & 0.00535 & -1.10 \\
calc. (6-$exp$) modif. & 0.030 & 0.134 & 0.010 & 0 & 36.6 & 3.00 & 0.0034 & -1.06 \\
exp. extrap. to 293 K & 0.12 & 0.06 & 0.03$\div $0.04 & 0 & 42$\div $45 & - & - & - \\ \hline
\end{tabular}
\end{ruledtabular}
\footnotetext{Note: see detailed explanation of "(6-$exp$) modif." parameterization in Subsection (\ref{cont}).}
\end{table*}

\begin{table*}
\caption{\label{bpzExpCalc}Comparison of experimental and calculated unit cell parameters for Fe(bpz)$_{2}$(bipy) (at 1 atm).}
\begin{ruledtabular}
\begin{tabular}{|l|c|c|c|c|c|c|c|c|}
\hline
& $a$, \AA & $b$, \AA & $c$, \AA & $\beta ,^{\circ }$ & $V$, \AA$^{3}$ & angle,$^{\circ }$ & CM $y/b$ & H, kcal/mol \\ \hline
\multicolumn{9}{|l|}{\textbf{The LS isomer}} \\ \hline
calc. (6-12) & 16.319 & 14.840 & 10.685 & 113.97 & 2364 & 92.55 & 0.2699 & -49.78 \\
calc. (6-$exp$) & 16.136 & 14.661 & 10.697 & 114.25 & 2307 & 91.93 & 0.2724 & -40.22 \\
exp. (139 K) \cite{BpzExp} & 16.086 & 14.855 & 10.812 & 114.18 & 2357 & 90.91 & 0.2754 & - \\ \hline
\multicolumn{9}{|l|}{\textbf{The HS isomer}} \\ \hline
calc. (6-12) & 16.242 & 15.178 & 10.823 & 113.60 & 2445 & 85.06 & 0.2703 & -48.59 \\
calc. (6-$exp$) & 16.032 & 14.995 & 10.834 & 113.92 & 2381 & 84.48 & 0.2728 & -39.75 \\
exp. (293 K) \cite{BpzExp} & 16.307 & 15.075 & 11.024 & 114.95 & 2457 & 85.02 & 0.2782 & - \\
exp. extrap. to 139 K & 16.16 & 14.99 & 11.04 & 114.9 & 2426$\div $2429 & - & - & - \\ \hline
\multicolumn{9}{|l|}{\textbf{The difference between the HS and LS isomers}} \\ \hline
calc. (6-12) & -0.077 & 0.338 & 0.138 & -0.37 & 81 & -7.49 & 0.0004 & 1.19 \\
calc. (6-$exp$) & -0.104 & 0.334 & 0.137 & -0.33 & 74 & -7.45 & 0.0004 & 0.47 \\
exp. extrap. to 139 K & 0.07 & 0.14 & 0.23 & 0.7 & 69$\div $72 & - & - & - \\
exp. (30 K) \cite{BpzLowT} & -0.076 & 0.347 & 0.219 & 1.09 & 71.2 & - & - & - \\ \hline
\end{tabular}
\end{ruledtabular}
\end{table*}

The changes of the unit cell volumes in the course of the spin transition are relatively small differences of two large numbers, and their correct estimation is difficult. For example, $\Delta V$ of [Fe(pyim)$_{2}$(bpy)](ClO$_{4}$)$_{2}\cdot $2C$_{2}$H$_{5}$OH was found to be negative \cite{Postnikov2008}, though all complexes studied experimentally have positive $\Delta V$, in agreement with the fact that the Fe$-$N bonds are longer in the HS complexes, and thus the HS molecules should have a larger "size". The calculated value of $\Delta V$ for [Fe(trim)$_{2}$]Cl$_{2}$, published in \cite{lemercier2006}, has the correct sign, but the experimental volumes of the LS and HS crystals are available only for different temperatures, which makes it impossible to compare the experimental and calculated values.

The values of $\Delta V$ of the Fe(phen)$_{2}$(NCS)$_{2}$ compound, calculated by us with both Lennard-Jones and Buckingham atom-atom potentials, are fairly close to the experimental values (extrapolated to the room temperature), being probably overestimated by $10-17\%$ (while the uncertainty in the extrapolated experimental value is \textit{ca.} $10$\%). In the case of Fe(btz)$_{2}$(NCS)$_{2}$, the errors are correspondingly about $+15\%$ and $-3\%$ for the two potentials, while the uncertainty in the extrapolated experimental value is \textit{ca.} $7\%$. Finally, in the case of Fe(bpz)$_{2}$(bipy) the calculated values differ from the experimental one, extrapolated to 139 K, by $4-14\%$. We would like to stress that the temperature dependence of $\Delta V$ is much stronger, than that of the unit cell volume $V$. For example, the low-temperature (at $15$ K) $\Delta V($Fe(phen)$_{2}$(NCS)$_{2})$ equals to $61.2$ \AA$^{3}$, the $\Delta V$ value extrapolated to 293 K is about 70$\div $79\AA$^{3}$, and the difference between the experimental unit cell volume of the HS form at $293$ K and that of the LS form at $130$ K is $119.1$ \AA$^{3}$. The presumable errors of the atom-atom potentials method in calculations of $\Delta V$ (\textit{ca.} $5-15$ \AA$^{3}$) are comparable with the uncertainties in the extrapolated estimates for experimental values (\textit{ca.} $3-9$ \AA$^{3}$) and much less than the changes in the volumes of the crystals caused by temperature expansion of the crystals (dozens of \AA$^{3}$). 

As for the \textit{unit cells} themselves, in all the cases the symmetry for the energy minimum points, according to our calculations, coincides with the experimental one. Orientation of a molecule in the unit cell can be characterized by three angles, corresponding to the transformation of coordinates from the molecular coordinate system (\textit{e.g.} that of the principal axes of inertia tensor) to the laboratory (or crystal) coordinate system. In all the considered cases, two of these angles have trivial values ($0$, $90$ or $180^{\circ }$); the values of the third angle, corresponding to rotation around the $C_{2}$ axis of the molecule, are given in Tables \ref{phenExpCalc}-\ref{bpzExpCalc}. The same is true for the CM positions of molecules within a unit cell. Two parameters out of three for each molecule are trivial ($0$, $1/4$, $1/2$, or $3/4$ of the corresponding translation period). The remaining parameter (corresponding to the $y$ coordinate in the units of $b$) is given in Tables \ref{phenExpCalc}-\ref{bpzExpCalc} as well. One can see that the calculated values are fairly close to the experimental ones both for the rotation angles and the CM positions.

The discrepancy between the calculated and experimental values of \textit{the lattice parameters} $a$, $b$, $c$ is in the $0.1$\% to $3.2$\% range, on average being equal to $1.3$\% for the (6-12) potential and $1.4$\% for the (6-$exp$) potential. As for the changes of these parameters in the course of the spin transition, in most cases the results predicted by the method of atom-atom potentials are in good agreement with the experimental data (the errors are \textit{ca.} $0.05-0.1$ \AA). What is especially impressing is that the method is capable of reproducing decrease of some lattice periods in the course of the spin transition, which may happen in spite of the overall increase of the unit cell volume (the parameter $b$ of the Fe(phen)$_{2}$(NCS)$_{2}$ crystal, the parameter $a$ of the Fe(bpz)$_{2}$(bipy) crystal). However, we have three problematic cases: the variation of the parameter $c$ of the Fe(phen)$_{2}$(NCS)$_{2}$ crystal (underestimated by the factor of 3$\div $5 times), and the variation of the parameters $a$ and $b$ of the Fe(btz)$_{2}$(NCS)$_{2}$ crystal (wrong sign of the result for $a$ and underestimation by the factor of 3$\div $4 times for $b$). It is especially important that both Lennard-Jones and Buckingham potentials yield close results. Trying to find an explanation for these errors, we noted that these three parameters are most sensitive to temperature changes. For example, the poorly predicted $\Delta c$ of the Fe(phen)$_{2}$(NCS)$_{2}$ crystal (calculated from the $c$ values extrapolated to the same temperature) changes in relative terms by $0.4\%$ per 100 K, while both $\Delta a$ and $\Delta b$ -- only by $0.2\%$ per 100 K. Similarly, $\Delta a$, $\Delta b$ and $\Delta c$ of the Fe(btz)$_{2}$(NCS)$_{2}$ crystal decrease by $0.7\%$, $1.1\%$ and $0.6\%$ per 100 K. This allows us to suggest that omission of the explicit treatment of the entropy contribution to the Gibbs energy, and thus the uncertainty in renormalization of the empirical parameters of the potentials, is one of the main sources of errors in the method in its current form, even if it is partially compensated by data correction for the thermal expansion.

Another possible explanation (which does not exclude the previous one) is that some specific interactions take place in these crystals, different from those in ordinary organic crystals used for fitting the presumably pure van der Waals interaction parameters. In this case, the performance of the method can be improved by correcting the parameters of atom-atom interactions.

\subsection{\label{cont}Contacts analysis and parameters adjustment}

\begin{figure}
\includegraphics[scale=0.44]{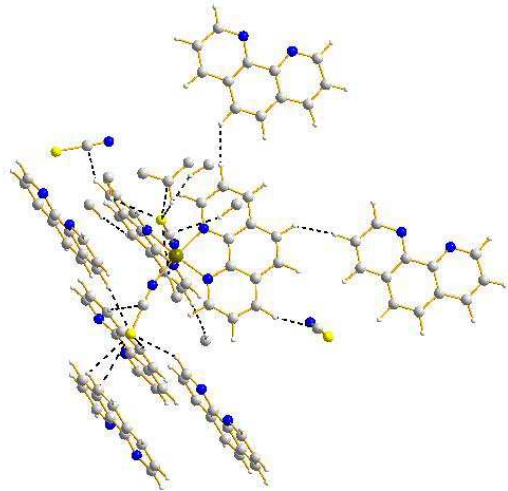}
\caption{\label{phen_cont}Contacts in the crystal of Fe(phen)$_{2}$(NCS)$_{2}$.}
\end{figure}

\begin{figure}
\includegraphics[scale=0.44]{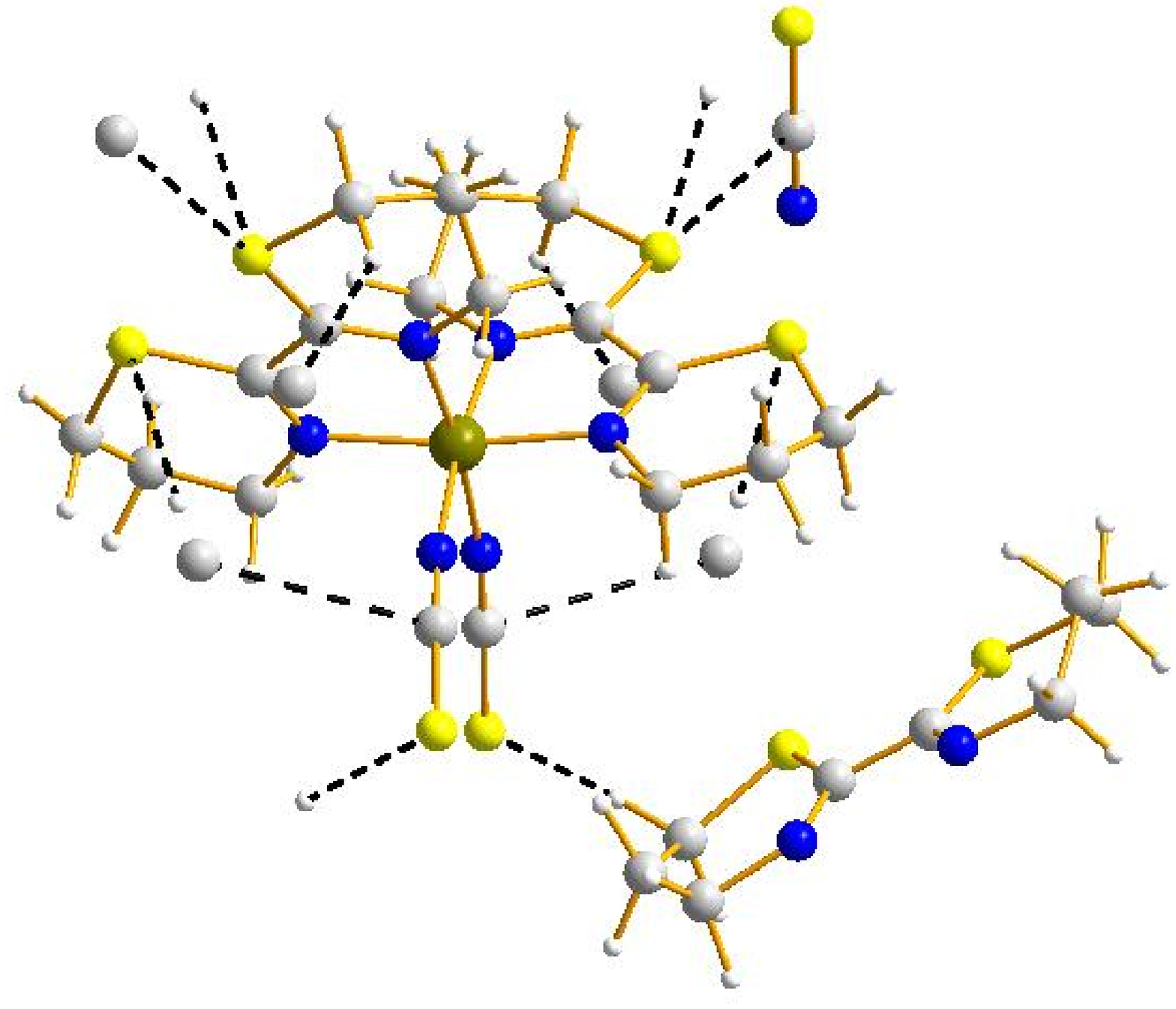}
\caption{\label{btz_cont}Contacts in the crystal of Fe(btz)$_{2}$(NCS)$_{2}$.}
\end{figure}

\begin{figure}
\includegraphics[scale=0.44]{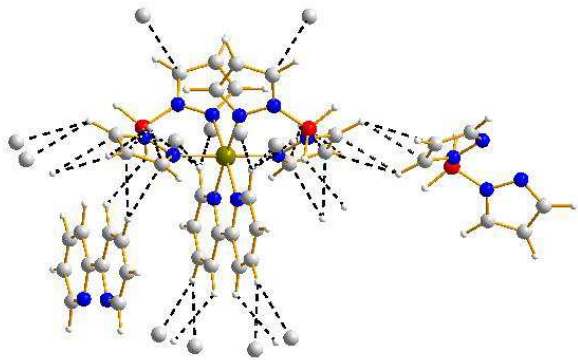}
\caption{\label{bpz_cont}Contacts in the crystal of Fe(bpz)$_{2}$(bipy).}
\end{figure}

To study this problem and yet further improve the performance of the method, we analyzed intermolecular contacts in the crystals, comparing atom-atom distances found in the experimental studies with those optimized with the parameters from \cite{UFFparam,Buckparam}. The lists of the shortest atom-atom contacts (we selected those separated by less than the sum of the corresponding van der Waals radii) are given in Tables \ref{phenContacts}-\ref{bpzContacts} and they are also depicted on Figs. \ref{phen_cont}-\ref{bpz_cont}. It is important to note that in all three materials the spin transition does not much affect the picture of intermolecular contacts. In other words, the shortest contacts in a LS crystal are also short (typically, though not always, shorter than the sum of the van der Waals radii) contacts in its HS form, and vice versa.

\begin{table}
\caption{\label{phenContacts}Shortest intermolecular contacts in the LS and HS crystals of Fe(phen)$_{2}$(NCS)$_{2}$: interatomic distances (\AA) determined experimentally (exp.) or calculated with (6-12) and (6-$exp$) potentials and the sum of the van der Waals radii (vdW) of the atoms \cite{Buckparam}.}
\begin{ruledtabular}
\begin{tabular}{|l|c|c|c|c|}
\hline
Pair & R(exp.) & R(vdW) & R(6-12) & R(6-$exp$) \\ \hline
\multicolumn{5}{|l|}{\textbf{The LS isomer}} \\ \hline
H$\cdots $H & 2.093 & 2.34 & 2.374 & 2.289 \\ \hline
S$\cdots $C & 3.314 & 3.55 & 3.370 & 3.341 \\ \hline
C$\cdots $H & 2.589, 2.784 & 2.92 & 2.620, 2.801 & 2.520, 2.866 \\ \hline
S$\cdots $H & 2.832, 2.891, & 2.97 & 3.162, 3.311, & 3.185, 3.419, \\
& 2.911, 2.951 & & 3.052, 2.907 & 2.983, 2.915 \\ \hline
\multicolumn{5}{|l|}{\textbf{The HS isomer}} \\ \hline
H$\cdots $H & 2.211 & 2.34 & 2.411 & 2.307 \\ \hline
S$\cdots $C & 3.357 & 3.55 & 3.345 & 3.339 \\ \hline
C$\cdots $H & 2.570, 2.750 & 2.92 & 2.635, 2.872 & 2.509, 2.792 \\ \hline
S$\cdots $H & 2.941 & 2.97 & 3.121 & 3.156 \\ \hline
\end{tabular}
\end{ruledtabular}
\end{table}

\begin{table}[tbp]
\caption{\label{btzContacts}Shortest intermolecular contacts in the LS and HS crystals of Fe(btz)$_{2}$(NCS)$_{2}$: interatomic distances (\AA) determined experimentally (exp.) or calculated with (6-12) and (6-$exp$) potentials and the sum of the van der Waals radii (vdW) of the atoms \cite{Buckparam}.}
\begin{ruledtabular}
\begin{tabular}{|l|c|c|c|c|}
\hline
Pair & R(exp.) & R(vdW) & R(6-12) & R(6-$exp$) \\ \hline
\multicolumn{5}{|l|}{\textbf{The LS isomer}} \\ \hline
S$\cdots $C & 3.275 & 3.55 & 3.352 & 3.333 \\ \hline
S$\cdots $H & 2.706, 2.743, & 2.97 & 2.886, 2.840, & 2.755, 2.741, \\
& 2.942 & & 2.849 & 2.907 \\ \hline
C$\cdots $H & 2.811 & 2.92 & 2.931 & 2.805 \\ \hline
C$\cdots $C & 3.453 & 3.50 & 3.668 & 3.570 \\ \hline
\multicolumn{5}{|l|}{\textbf{The HS isomer}} \\ \hline
S$\cdots $C & 3.351 & 3.55 & 3.401 & 3.384 \\ \hline
S$\cdots $H & 2.893, 2.925 & 2.97 & 2.832, 2.904 & 2.770, 2.805 \\ \hline
C$\cdots $H & 2.848, 2.888 & 2.92 & 2.802, 2.851 & 2.693, 2.755 \\ \hline
\end{tabular}
\end{ruledtabular}
\end{table}

\begin{table}[tbp]
\caption{\label{bpzContacts}Shortest intermolecular contacts in the LS and HS crystals of Fe(bpz)$_{2}$(bipy): interatomic distances (\AA) determined experimentally (exp.) or calculated with (6-12) and (6-$exp$) potentials and the sum of the van der Waals radii (vdW) of the atoms \cite{Buckparam}.}
\begin{ruledtabular}
\begin{tabular}{|l|c|c|c|c|}
\hline
Pair & R(exp.) & R(vdW) & R(6-12) & R(6-$exp$) \\ \hline
\multicolumn{5}{|l|}{\textbf{The LS isomer}} \\ \hline
C$\cdots $H & 2.655, 2.658, & 2.92 & 2.732, 2.716, & 2.651, 2.606, \\
& 2.689, 2.817, & & 2.815, 2.877, & 2.720, 2.763, \\
& 2.879, 2.912 & & 2.895, 2.811 & 2.835, 2.818 \\ \hline
H$\cdots $H & 2.283, 2.332 & 2.34 & 2.387, 2.390 & 2.301, 2.333 \\ \hline
C$\cdots $C & 3.368, 3.374 & 3.50 & 3.327, 2.720 & 3.387, 3.412 \\ \hline
\multicolumn{5}{|l|}{\textbf{The HS isomer}} \\ \hline
C$\cdots $H & 2.579, 2.719, & 2.92 & 2.728, 2.739, & 2.622, 2.657, \\
& 2.782, 2.813, & & 3.002, 2.809, & 2.892, 2.707 \\
& 2.878, 2.900 & & 2.838, 3.189 & 2.769, 3.040 \\ \hline
H$\cdots $H & 2.318 & 2.34 & 2.477 & 2.336 \\ \hline
C$\cdots $C & 3.360, 3.420 & 3.50 & 3.250, 3.304 & 3.194, 3.236 \\ \hline
N$\cdots $H & 2.606 & 2.67 & 2.786 & 2.642 \\ \hline
\end{tabular}
\end{ruledtabular}
\end{table}

First of all, one can see that in most cases the shortest contacts involve hydrogen atoms (S$\cdots $H,
C$\cdots $H, N$\cdots $H, or H$\cdots $H). It is well known that coordinates of the hydrogen atoms
determined from X-ray diffraction may be subject to significant errors unless tricks of crystallographic
computing are used. While the X--H bond length is notoriously underestimated due to the shift of the 
bonding electron pair towards the nonmetal X atom, the position of the X--H vector in three-dimensional
space is correctly found. Thus, the H atom should "ride" on the nonmetal atom with a {\em fixed} bond length
(e.g., C--H = 1.09 \AA, N--H = 1.01 \AA, O--H = 0.96 \AA). Because the crystal structures under study 
seemingly did not profit from such "riding" H atoms approach, we may suggest that one of the
main sources of mistakes in our results is the uncertainty in the H positions. This also
indicates that in the future research, when taking into consideration intramolecular degrees of freedom,
one should take possible deformations of the C$-$H bonds into account.

It is reasonable to suggest that the poorly described atom-atom interactions will be at the top of the list of the highest atom-atom repulsion energies. Indeed, if some interatomic distance increases when the system goes from the experimental configuration to the optimized one, the repulsion between the corresponding atoms weakens. Thus one can expect that the intensity of that interaction is overestimated, since such relaxation does not occur in experiment. A similar reasoning applies to the strongest attractions as well. In practice, the picture is not so clear because molecules in organic crystals typically have numerous contacts between various atoms. By analyzing the crystals formed by the Fe(phen)$_{2}$(NCS)$_{2}$ or Fe(btz)$_{2}$(NCS)$_{2}$ molecules we found that the sulphur atoms play very important role in the intermolecular interactions (Fe(bpz)$_{2}$(bipy) does not contain sulphur). As our calculations demonstrated, the S atoms participate in many close contacts with other atoms, thus providing a significant contribution to the \textit{repulsion} within crystals; at the same time, their contributions to the \textit{attraction} are also dominant (attraction energies of various S$\cdots $S pairs are the largest by absolute value in these crystals; as for the S$\cdots $C contacts, in some of them attraction is also very strong, while some other S$\cdots $C contacts are among extreme cases of repulsion).

The Fe(phen)$_{2}$(NCS)$_{2}$ molecule has S atoms only in the NCS groups while in the case of Fe(btz)$_{2}$(NCS)$_{2}$ the chelating ligand also contains the S atoms. We found that the S atoms of both types participate in the contacts with extremal values of the energy. Taking into consideration that the parameterization of the van der Waals energy of the S$\cdots $X contacts (X = S, C, H) is not so well studied as compared to the C$\cdots $C, C$\cdots $H, and H$\cdots $H interactions, and that some involvement of the lone pairs and vacant $d$-orbitals of the S atoms can complicate the approximation of the S$\cdots $X interactions by the center-symmetric atom-atom contributions, we suggest that improving the treatment of S$\cdots $X (X = S, C, H) interaction energies may be another way of developing a better model of the atom-atom potentials for molecular crystals undergoing spin transitions. For example, the shortest C$\cdots $S distances are found to be \textit{ca.} $0.2$ \AA\ shorter than the sum of the van der Waals radii of the atoms. In the case of the S$\cdots $H contacts, this contraction may reach even $0.27$ \AA. Thus it is reasonable to suggest that due to some specific interactions, the optimal interatomic distances involving S atoms may be lower than determined by the standard parameterization.

For that reason we adjusted the parameters for the S$\cdots $C, S$\cdots $S, S$\cdots $H interactions in order to improve agreement between the experimental and modeled crystal lattice parameters. However, improvement of some of the calculated lattice parameters often increases the discrepancies for others. The situation is especially difficult for the differences between the spin isomers $\Delta a$, $\Delta b$, $\Delta c$. Variation of the parameters for atom-atom contacts similarly affects the lattice parameters in the LS and HS crystals, hence the resulting change in those parameters is small and it can be only calculated rather than predicted from any physical or geometrical reasoning.

We performed a systematic quantitative study of the influence of the interaction parameters on the equilibrium configurations of the crystals. To get the general understanding of this issue, we optimized the crystals of Fe(btz)$_{2}$(NCS)$_{2}$ with the interaction parameters slightly modified. We increased, one by one, parameters for each pair of atoms (the well depth and the equilibrium separation) by $5$\% to estimate numerically the sensitivity of the energy contributions to the potential parameters. The choice of the Fe(btz)$_{2}$(NCS)$_{2}$ crystals was suggested by the fact that it is poorly described with the original parameterization: there are qualitative discrepancies for the changes in two out of three lattice parameters ($\Delta a$ and $\Delta b$). Also we limited the consideration to the (6-$exp$) potential only, since the parameters of interaction between atoms of different elements were determined explicitly \cite{Buckparam} without any reference to the superposition approximation (except for the N$\cdots $S pairs making little difference for the systems studied), and thus they can be varied separately.
%calculated in the superpositional approximation. It would be
%methodologically correct in this case to make independent variations only in
%the parameters for homogeneous pairs of atoms, automatically adjusting the
%corresponding heterogeneous parameters, which would be not so informative.

\begin{table*}
\caption{\label{fittingParamsBtz}Changes in the optimal lattice parameters $a$, $b$ and $c$ (10$^{-4}$\AA) of Fe(btz)$_{2}$(NCS)$_{2}$ caused by 5\% increase of the (6-$exp$) potential parameters.}
\begin{ruledtabular}
\begin{tabular}{|l|c|c|c|c|c|c|c|c|c|}
\hline
& \multicolumn{3}{c|}{The LS isomer} & \multicolumn{3}{c|}{The HS isomer} & \multicolumn{3}{c|}{The HS/LS difference} \\ \hline
Pair & $\delta a$ & $\delta b$ & $\delta c$ & $\delta a$ & $\delta b$ & $\delta c$ & $\delta \Delta a$ & $\delta \Delta b$ & $\delta \Delta c$ \\ \hline
\multicolumn{10}{|c|}{well depth} \\ \hline
H$\cdots $H & 12 & 14 & -3 & 28 & -1 & -5 & 16 & -15 & -2 \\ \hline
C$\cdots $H & -14 & -11 & 1 & 6 & -9 & 11 & 20 & 2 & 10 \\ \hline
N$\cdots $H & -13 & -5 & -13 & -15 & -6 & -13 & -2 & -1 & 0 \\ \hline
S$\cdots $H & 126 & 24 & 6 & 85 & 41 & -10 & -41 & 17 & -16 \\ \hline
C$\cdots $C & -28 & -20 & -19 & -27 & -20 & -14 & 1 & 0 & 5 \\ \hline
N$\cdots $C & -25 & -10 & -31 & -29 & -12 & -27 & -4 & -2 & 4 \\ \hline
S$\cdots $C & -25 & 27 & 51 & -11 & 18 & 31 & 14 & -9 & -20 \\ \hline
N$\cdots $N & -11 & -4 & -1 & -13 & -4 & 3 & -2 & 0 & 4 \\ \hline
S$\cdots $S & -23 & -13 & 4 & -26 & -8 & 19 & -3 & 5 & 15 \\ \hline
\multicolumn{10}{|c|}{equilibrium distance} \\ \hline
H$\cdots $H & 251 & 216 & 5 & 483 & 24 & -33 & 232 & -192 & -38 \\ \hline
C$\cdots $H & 396 & 98 & 365 & 755 & 74 & 421 & 359 & -24 & 56 \\ \hline
N$\cdots $H & -60 & -29 & -48 & -41 & -23 & -21 & 19 & 6 & 27 \\ \hline
S$\cdots $H & 2084 & 828 & 700 & 1741 & 991 & 533 & -343 & 163 & -167 \\ \hline
C$\cdots $C & 168 & -55 & 269 & 277 & -61 & 263 & 109 & -6 & -6 \\ \hline
N$\cdots $C & -148 & -83 & -70 & -137 & -90 & -2 & 11 & -7 & 68 \\ \hline
S$\cdots $C & 714 & 1131 & 2075 & 894 & 958 & 1821 & 180 & -173 & -254 \\ \hline
N$\cdots $N & -80 & -27 & 271 & -109 & -18 & 305 & -29 & 9 & 34 \\ \hline
S$\cdots $S & 11 & -58 & 1116 & -22 & 28 & 1226 & -33 & 86 & 110 \\ \hline
exp. & 705 & 2930 & 1354 & 2098 & 1843 & 1131 & 1393 & -1087 & -223 \\ \hline
\end{tabular}
\end{ruledtabular}
\end{table*}

Changes in the optimal lattice parameters $\delta a$, $\delta b$, $\delta c$, caused by $5$\% variations of each parameter of the atom-atom interaction energy, are given in Table \ref{fittingParamsBtz}. The table also specifies the differences between the experimental values of the lattice parameters and those calculated with the initial parameters of \cite{Buckparam} (exp.). As one can see from the numbers, most of the interaction parameters very slightly affect the optimal configuration of the crystals. Corrections caused by \textit{the well depth} changes by $5$\% are a hundred times smaller than the difference between the experimental and calculated lattice parameters, leaving no hope to reduce the discrepancy by fitting the well depths within reasonable frames. The same applies to most of the \textit{atom-atom equilibrium separations} on the corresponding interaction energy curves (C$\cdots $C, H$\cdots $H, N$\cdots $N, N$\cdots $H, \textit{etc.}), though in this case the changes in the lattice parameters caused by a $5$\% increase of the distances are only tens times smaller than the required scale of correction.

Only three parameters significantly affect the optimal structure of the crystal: the equilibrium separations for the S$\cdots $C, S$\cdots $H, and S$\cdots $S pairs (listed in the order of decreasing effect). This confirms our assumption made above on the basis of the interatomic contacts analysis that the contacts involving the S atoms need an improved treatment first of all.

To do this, we performed a numerical minimization of the sum of squares of residuals $f$ as a function of the equilibrium separations $r$ for the S$\cdots $C, S$\cdots $H, and S$\cdots $S atoms:
\begin{equation}
\begin{array}{c}
f=
\left( a_{LS,calc}-a_{LS,exp} \right)^2+\\
\left( b_{LS,calc}-b_{LS,exp} \right)^2+
\left( c_{LS,calc}-c_{LS,exp} \right)^2+\\
\left( a_{HS,calc}-a_{HS,exp} \right)^2+
\left( b_{HS,calc}-b_{HS,exp} \right)^2+\\
\left( c_{HS,calc}-c_{HS,exp} \right)^2+
\left( \Delta a_{calc}-\Delta a_{exp} \right)^2+\\
\left( \Delta b_{calc}-\Delta b_{exp} \right)^2+
\left( \Delta c_{calc}-\Delta c_{exp} \right)^2,
\end{array}
\label{lsq}
\end{equation}
where $a_{calc}$, $b_{calc}$, $c_{calc}$ stand for the optimal lattice parameters calculated with the Buckingham potential parameterization different from one in \cite{Buckparam} by $r_{SC}$, $r_{SH}$ and $r_{SS}$ separations, and $a_{exp}$, $b_{exp}$, $c_{exp}$ are the experimental lattice parameters (extrapolated to 293 K, if necessary). The result is that the equilibrium separation of the S$\cdots $C contact should be increased by $6.6$\% ($+0.26$ \AA), that of the S$\cdots $H one -- decreased by $2.3$\% ($-0.08$ \AA), and that of the S$\cdots $S one -- decreased by $15$\% ($-0.57$ \AA). Optimal values of the crystal lattice parameters of the Fe(btz)$_{2}$(NCS)$_{2}$ crystal, calculated with the adjusted parameters, are given in Table \ref{btzExpCalc} (calc. (6-$exp$) modif.).

After fitting the S$\cdots $C, S$\cdots $H and S$\cdots $S equilibrium distances all six unit cell dimensions became closer to the experimental values: the error in $b$(LS) decreased from $-0.32$ \AA\ to $-0.15$ \AA, the error in $a$(HS) -- from $-0.21$ \AA\ to $-0.10$ \AA, in $b$(HS) -- from $-0.18$ \AA\ to $-0.08$ \AA, and so on. The values of the CM position and the rotation angle change, by contrast, insignificantly, in spite of the fact that they were not included in the treatment by the least squares method. The performance of the model in predicting the quantities $\Delta a$, $\Delta b$ and $\Delta c$ also significantly improved. The value of $\Delta a$ shifted towards the experimental one and changed the sign to the correct one (positive instead of negative). $\Delta b$ moved towards the experimental value, though this correction was only 1/3 of the initial discrepancy. Finally, the $\Delta c$ value shifted in the correct direction, but this time the change was even larger than the required one.  An attempt to improve yet further the relation between the predicted and actual values of $\Delta a$, $\Delta b$, $\Delta c$ and $\Delta V$ by another modification of the atom-atom interaction parameters (for example, by increasing the weights ascribed to the corresponding squares in the treatment by the least squares method) leads to catastrophic results for $a$, $b$ and $c$ of the pure LS and HS crystals: a tiny improvement by $0.01$ \AA\ in $\Delta a$, $\Delta b$, $\Delta c$ simultaneously leads to the growth in the discrepancies in $a$, $b$ and $c$ by \textit{ca.} $0.1$ \AA.

We applied the same modified parameterization to the LS and HS crystals of Fe(phen)$_{2}$(NCS)$_{2}$. The results are given in Table \ref{phenExpCalc} (calc. (6-$exp$) modif.). In regard to the unit cell dimensions of the LS and HS crystals, the modification improved 4 out of 6 periods, especially those poorly described by the original parameterization: the error in $b$(LS) decreased from $-0.32$ \AA\ to $-0.19$ \AA, in $b$(HS) -- from $-0.29$ \AA\ to $-0.15$ \AA. Error in unit cell volumes $V$ decreased 2-3 times. At the same time, a significant error in the $c$ value for the HS form appeared ($0.33$ \AA\ instead of $0.06$ \AA). As for the changes in the lattice parameters, their values became more distant from the experimental values by $0.01-0.09$ \AA. To sum up, the suggested modification generally improves the results of the model for both S-containing materials, though it fails to eliminate the errors completely.

The adjustment of interaction parameters, described in this Subsection, does not claim to produce a new system of atom-atom parameters. We undertook it just to estimate how much improvement in the performance of the method at the expense of minor changes within the same theoretical paradigm may be done, and to illustrate that accurate treatment of the intermolecular contacts involving sulphur atoms are of primary importance for modeling the spin transition in S-containing materials.

\subsection{\label{enth}Contributions of intermolecular interactions to enthalpy}
The results described in the previous Sections demonstrate that the method of atom-atom potentials is capable of modeling intermolecular interactions and reproducing experimental data on the geometry of the unit cells. This allows us to go on to estimate the contributions of the van der Waals intermolecular forces to the energy (enthalpy) of the spin transitions, which cannot be extracted from experimental data. The results are given in the last columns of Tables \ref{phenExpCalc}-\ref{bpzExpCalc}. First of all, one can see that this contribution may be either positive or negative, which corroborates the theoretical conclusion of \cite{TDIJQC}. Another important point is that the lattice contribution to the enthalpy of the spin transition is comparable with its total value. Though the estimates obtained with the Lennard-Jones and Buckingham potentials are somewhat different, the general picture is the same. For example, in the case of the Fe(phen)$_{2}$(NCS)$_{2}$ crystal we found this component to be equal to $+1.81$ kcal/mol (6-12) or $+0.57$ kcal/mol (6-$exp$) or $+1.54$ kcal/mol (6-$exp$ modified), while the total experimental enthalpy (from the calorimetrical data) is $+2.05$ kcal/mol \cite{SoraiSeki}. It means that one cannot neglect intermolecular interactions in calculating thermodynamical characteristics of the spin transitions in molecular crystals. (This conclusion was also made in \cite{lemercier2006} on the basis of DFT calculations; however, the contribution of intermolecular interactions, which can be extracted from their results and ranging from $2$ to $23$ kcal/mol, seems to be strongly overestimated).

\begin{table}
\caption{\label{phenHighP}Comparison of experimental and calculated unit cell parameters for Fe(phen)$_{2}$(NCS)$_{2}$ (at 1 GPa).}
\begin{ruledtabular}
\begin{tabular}{|l|l|c|c|c|c|}
\hline
System &  & $a$, \AA & $b$, \AA & $c$, \AA & $V$, \AA$^{3}$ \\ \hline
LS(1 GPa) & calc. (6-12) & 13.060 & 9.773 & 17.183 & 2193.2 \\
& calc. (6-$exp$) & 12.838 & 9.700 & 17.089 & 2128.0 \\
& exp. (298 K) \cite{phenHighP} & 12.656 & 9.848 & 16.597 & 2068.6 \\ \hline
Difference, & calc. (6-12) & -0.465 & -0.137 & -0.399 & -163.5 \\
LS(1 GPa)/ & calc. (6-$exp$) & -0.426 & -0.169 & -0.453 & -168.2 \\
HS(1 atm) & exp. & -0.505 & -0.315 & -0.884 & -269.6 \\ \hline
Difference, & calc. (6-12) & -0.125 & -0.149 & -0.163 & -76.0 \\
LS(1 GPa)/ & calc. (6-$exp$) & -0.155 & -0.161 & -0.193 & -86.0 \\
LS(1 atm) & exp. & -0.114 & -0.242 & -0.625 & -150.5 \\ \hline
\end{tabular}
\end{ruledtabular}
\end{table}

\begin{table}
\caption{\label{btzHighP}Comparison of experimental and calculated unit cell parameters for Fe(btz)$_{2}$(NCS)$_{2}$ (at 1 GPa).}
\begin{ruledtabular}
\begin{tabular}{|l|l|c|c|c|c|}
\hline
System &  & $a$, \AA & $b$, \AA & $c$, \AA & $V$, \AA$^{3}$ \\ \hline
LS(1 GPa) & calc. (6-12) & 13.072 & 10.410 & 16.640 & 2264.4 \\
& calc. (6-$exp$) & 12.877 & 10.380 & 16.502 & 2205.7 \\
& exp. (298 K) \cite{phenHighP} & 12.839 & 10.454 & 16.362 & 2196.1 \\ \hline
Difference, & calc. (6-12) & -0.171 & -0.313 & -0.307 & -142.2 \\
LS(1 GPa)/ & calc. (6-$exp$) & -0.366 & -0.343 & -0.445 & -200.9 \\
HS(1 atm) & exp. & -0.449 & -0.407 & -0.558 & -245.8 \\ \hline
Difference, & calc. (6-12) & -0.194 & -0.107 & -0.249 & -92.0 \\
LS(1 GPa)/ & calc. (6-$exp$) & -0.389 & -0.138 & -0.386 & -150.7 \\
LS(1 atm) & exp. & -0.216 & -0.196 & -0.310 & -122.0 \\ \hline
\end{tabular}
\end{ruledtabular}
\end{table}

\subsection{\label{press}Pressure effects}

Finally, we studied behavior of the crystal lattice parameters under the external hydrostatic pressure. Calculations were made for the Fe(phen)$_{2}$(NCS)$_{2}$ and Fe(btz)$_{2}$(NCS)$_{2}$ compounds, since the experimental data on the pressure effects on the spin transition are available only for these crystals \cite{phenHighP}. We performed minimization of the enthalpies as a function of the lattice parameters, CM positions of the molecules, and their rotation angles, at two values of the pressure. The external pressure was accounted for by the $PV$ term in the function to be minimized. The starting points of optimization were the experimental geometries. As previously, at the first step we minimized enthalpy as a function of five parameters, preserving the symmetry of the crystal, and after that checked that we get the global minima by allowing variation of all 27 parameters mentioned above. The results for the lattice parameters of the LS forms at 1 GPa and $298$ K are given in Tables \ref{phenHighP}, \ref{btzHighP}, and the compressibility coefficients at 1 atm and 1 GPa -- in Table \ref{compressCoeff}. As one can see from the tables, the high-pressure lattice parameters are very well reproduced (errors are below $4$\%), though less accurately than those for the low pressure. As for the compressibility coefficients, in all the cases the correspondence between the calculated and experimental values is qualitative (the compressibility coefficients are underestimated by a factor of 1.5$\div $3 as compared to the experimental values). One can see that the Buckingham potential produces better values than the Lennard-Jones one. Taking into consideration that the (6-$exp$) parameterization used in the present study is based only on the crystal structures measured at 1 atm, and very few contacts in those structures have distances corresponding to the repulsive branch of the potential (see Figs. 5-7 of Ref. \cite{Buckparam}), we conclude that our results for the high-pressure structures are better than one could expect.

\begin{table}
\caption{\label{compressCoeff}Comparison of experimental and calculated compressibility coefficients (in $10^{-1}$ GPa$^{-1}$) for Fe(phen)$_{2}$(NCS)$_{2}$ and Fe(btz)$_{2}$(NCS)$_{2}$.}
\begin{ruledtabular}
\begin{tabular}{|l|l|c|c|c|c|}
\hline
System &  & $k_{a}$ & $k_{b}$ & $k_{c}$ & $k_{V}$ \\ \hline
Fe(phen)$_{2}$(NCS)$_{2}$ & calc. (6-12) & 0.14 & 0.22 & 0.14 & 0.50 \\
HS, 1 atm & calc. (6-$exp$) & 0.16 & 0.22 & 0.17 & 0.56 \\
& exp. (298 K) \cite{phenHighP} & 0.21 & 0.33 & 0.53 & 1.07 \\ \hline
Fe(phen)$_{2}$(NCS)$_{2}$ & calc. (6-12) & 0.07 & 0.12 & 0.07 & 0.26 \\
LS, 1 GPa & calc. (6-$exp$) & 0.09 & 0.13 & 0.09 & 0.30 \\
& exp. (298 K) \cite{phenHighP} & 0.16 & 0.28 & 0.38 & 0.82 \\ \hline
Fe(btz)$_{2}$(NCS)$_{2}$ & calc. (6-12) & 0.22 & 0.14 & 0.20 & 0.56 \\
HS, 1 atm & calc. (6-$exp$) & 0.25 & 0.14 & 0.18 & 0.57 \\
& exp. (298 K) \cite{phenHighP} & 0.41 & 0.43 & 0.37 & 1.21 \\ \hline
Fe(btz)$_{2}$(NCS)$_{2}$ & calc. (6-12) & 0.11 & 0.08 & 0.11 & 0.30 \\
LS, 1 GPa & calc. (6-$exp$) & 0.13 & 0.09 & 0.11 & 0.34 \\
& exp. (298 K) \cite{phenHighP} & 0.28 & 0.33 & 0.28 & 0.89 \\ \hline
\end{tabular}
\end{ruledtabular}
\end{table}

\section{\label{concl}Conclusion}

Numerical modeling of the spin transitions in molecular crystals is important from practical and theoretical viewpoints. There is no alternative to calculations explicitly taking into account the composition and structure of interacting molecules (instead of representing them by spheres, or ellipsoids, or octahedra \textit{etc.}, immersed in an elastic media), both for the purposes of theoretical study of the transition mechanisms and for prediction of phenomenological parameters for macroscopic models. Meanwhile, the modern quantum chemical methods are hardly applicable to such objects, because their accuracy level is not sufficient to calculate the required values (for example, enthalpies of the spin transitions).

We demonstrate that the atom-atom potentials can be used for analysis of intermolecular contributions to the structure and energy of spin-active crystals. Indeed, intermolecular contacts in these crystals are those between the C, H, N, S, \textit{etc}. organogenic atoms, while the metal atom and its bonds with the donor atoms of the ligands are hidden inside the complex. As a consequence of that, the van der Waals interactions in the spin-active crystals can be approximated similarly to those in ordinary organic molecular crystals. In the present paper we checked for the first time the possibility to use the atom-atom potentials method for this class of objects.

In all the cases the symmetry groups of optimized crystals coincided with those found in experiment; the unit cell volumes were calculated with the precision of $0.5-4$\%. Errors in the predicted lattice parameters did not exceed $3$\% at the ambient pressure and $4$\% at 1 GPa. Direction (sign) and magnitude of the changes of the lattice parameters and molecules positions in the unit cell in the course of the temperature- and pressure-driven spin transitions were reproduced correctly. The compressibility coefficients are in a qualitative agreement with their experimental values, although 1.5$\div $3 times underestimated. Thus the accuracy of the method of atom-atom potentials is quite sufficient at the present level of the theory. We attempted to improve the parameterization, which is based on the (6-$exp$) parameters from \cite{Buckparam} and differs from it in the S$\cdots $H, S$\cdots $C and S$\cdots $S equilibrium separations. The results of this fitting of the parameters demonstrate that the performance of the method can be significantly improved by adjustment to the specific cases under study, and that the energy of interactions involving sulphur atoms is the crucial term for adequate treatment of spin transitions in the crystals studied.

Our study shows that any reliable calculation of spin transition parameters (such as transition enthalpy) must take into account intermolecular interactions. According to our estimates for the Fe(phen)$_{2}$(NCS)$_{2}$ crystal, the van der Waals contribution to the transition enthalpy is about +0.6$\div $1.8 kcal/mol (as compared with the total transition enthalpy of $+2.05$ kcal/mol).

We believe that the accuracy of the method used in this paper is limited by (i) implicit treatment of the entropy effects (through fitting the interaction parameters, rather than explicit calculation of frequencies of intermolecular oscillations); (ii) uncertainty of the H atoms positions in the experimental X-ray structures; (iii) description of the energy of interactions involving the S atoms (due to possible involvement of lone pairs and vacant $d$-orbitals of the sulphur atoms). Nevertheless, even the current level of precision is enough for using the method of atom-atom potentials to study the spin transitions in molecular crystals.
%At any rate, accuracy of the crystal structure parameters predicted by the
%atom-atom potentials method is significantly better than one of the published DFT
%estimates. In the meantime, computational facilities required for it are
%incomparably smaller.

\section*{Acknowledgments}

This work has been supported by the RFBR through the grant No 07-03-01128. The financial support of this work through the JARA-SIM research project "Local Electron States in Molecules and Solids" is gratefully acknowledged. The authors are thankful to Prof. J. \'Angy\'an of Universit\'e Henri Poincar\'e, Nancy for valuable discussion and sending his results prior to publication. Valuable discussions with Prof. A.V. Yatsenko and Dr. N.V. Goulioukina of the Chemistry Department of Moscow State University are gratefully acknowledged.

\end{document}